\documentclass[twocolumn,preprintnumbers,amsmath,amssymb]{revtex4}
\usepackage{amssymb}
\usepackage{latexsym}
\usepackage{epsfig}
\usepackage{color}

\begin{document}
\title{Generalized entropy formalism and a new holographic dark energy model}
\author{A. Sayahian Jahromi$^1$, S. A. Moosavi$^2$, H. Moradpour$^2$\footnote{h.moradpour@riaam.ac.ir},
J. P. Morais Gra\c ca$^3$, I. P. Lobo$^3$, I. G. Salako$^4$, A.
Jawad$^5$}
\address{$^1$ Zarghan Branch, Islamic Azad University, Zarghan, Iran\\
$^2$ Research Institute for Astronomy and Astrophysics of Maragha
(RIAAM), P.O. Box 55134-441, Maragha, Iran\\
$^3$ Departamento de F\'{i}sica, Universidade Federal da
Para\'{i}ba, Caixa Postal 5008, CEP 58051-970, Jo\~{a}o Pessoa,
PB, Brazil\\
$^4$ Institut de Math\'ematiques et de Sciences Physiques (IMSP)
01 BP 613 Porto-Novo, B\'enin\\
$^5$ Department of Mathematics, COMSATS Institute of\\
Information Technology, Lahore-54000, Pakistan}
\begin{abstract}
Recently, the R\'{e}nyi and Tsallis generalized entropies have
extensively been used in order to study various cosmological and
gravitational setups. Here, using a special type of generalized
entropy, a generalization of both the R\'{e}nyi and Tsallis
entropy, together with holographic principle, we build a new model
for holographic dark energy. Thereinafter, considering a flat FRW
universe, filled by a pressureless component and the new obtained
dark energy model, the evolution of cosmos has been investigated
showing satisfactory results and behavior. In our model, the
Hubble horizon plays the role of IR cutoff, and there is no mutual
interaction between the cosmos components. Our results indicate
that the generalized entropy formalism may open a new window to
become more familiar with the nature of spacetime and its
properties.
\end{abstract}

\maketitle

\section{Introduction\label{Intr}}

In standard cosmology, based on general relativity, one way to
describe the current accelerating universe is to consider an
unknown energy-momentum source called dark energy
\cite{Roos,Rev1,Rev2}. From thermodynamic point of view, dark
energy candidates and horizon entropy can be affected by each
other \cite{mswr,em,DEC37,ijtpmr,pav}. Recently, due to the
unknown nature of spacetime, the long-range nature of gravity, and
also motivated by the fact that the Bekenstein-Hawking entropy is
a non-extensive entropy measure \cite{nn1,nn2,non2,non3,5}, the
R\'{e}nyi and Tsallis generalized entropies
\cite{nn1,nn2,nn3,non0,non1,abe,1,2,fon,5} have been attributed to
horizons to study various cosmological and gravitational phenomena
\cite{non2,non15,non16,non17,non18,non19,non20,non21,non22,non23,
non3,non13,non4,non5,non6,non7,non8,non9,non10,non11,non12,non14,eb,eb1,3,4,5,6,7,8,9,10,11}.
The successes of these attempts in modelling the current
accelerating cosmos
\cite{non4,non5,non6,non7,non8,non9,non10,non14,non13,non19,non20,
non21} encourage and motivate us to study the cosmos evolution in
various generalized entropy setups which may help us to become
familiar with the probable non-extensive features of spacetime,
and thus its origin \cite{non20}.

Based on spacetime thermodynamics, the apparent horizon of FRW
universe is a proper causal boundary \cite{Hay2,Hay22,Bak},
meaning that the thermodynamics laws are satisfied on this
boundary \cite{Cai2,CaiKim}. Moreover, WMAP data indicates a flat
FRW universe \cite{Roos}, a universe for which apparent horizon is
equal to the Hubble horizon. Thus, proper models of dark energy
should be in agreement with the Hubble horizon in flat FRW background.

Following the Cohen et al's hypothesis on the mutual relation
between the UV cutoff and the entropy of system \cite{HDE}, a new
class of dark energy models have been proposed, called holographic
dark energy (HDE)
\cite{HDE01,HDE1,HDE2,HDE3,HDE4,HDES,HDE5,HDE22,RevH,wang}. In
flat FRW universe, the original model of HDE (OHDE) is constructed
by attributing the Bekenstein-Hawking entropy to the cosmos
horizon and also considering the Hubble horizon as its IR cutoff
\cite{HDE01,HDE1,HDE2,HDE3,HDE4,HDES}. Although the density
parameter of OHDE shows an admissible behavior from itself, its
energy density scales with $H^2$ meaning that it behaves as dark
matter during the cosmos evolution \cite{HDE3,HDE4}, and in fact,
OHDE is not in harmony with the Hubble radius \cite{HDE3,HDE4}.
Besides, it is not always stable whenever it is dominant in cosmos
and controls its expansion rate \cite{HDES}. Due to such
weaknesses of OHDE, various attempts have been made to modify this
model \cite{RevH,wang}.

Although various entropies have been used to get modified HDE
\cite{RevH,wang}, none of them consider the generalized entropy
formalism to build a HDE model. As we have previously mentioned,
the R\'{e}nyi and Tsallis generalized entropies generate suitable
models for the current universe, and thus, we are going to use
such formalism to build a new model for HDE in flat FRW by
considering the Hubble radius as its IR cutoff. Here, we use a
special generalized entropy, a generalization of both the
R\'{e}nyi and Tsallis entropies, to build our model. In fact, our
final aim of introducing this new holographic model is to show that
the probable non-additive and non-extensive aspects of spacetime have
theoretically enough potential to accelerate the universe in a
consistent way with observations.

The paper is organized as follows. In the next section, after
reviewing some generalized entropy formalisms, we introduce our
model of HDE. In continue, we consider a non-interacting universe,
for which there is no mutual interaction between the cosmos
components, and study the evolution of system in
Sec.~($\textmd{III}$). A summary on the present work is also
presented in the last section. The unit of $c=\hbar=G=k_B=1$,
where $k_B$ denotes the Boltzmann constant, has also been used in
this paper.


\section{horizon entropy in generalized entropy formalism and holographic dark energy}

Consider a system including $W$ states, in which $P_i$ is the
probability of achieving the $i^{th}$ state satisfying the
$\sum_{i=1}^W P_i=1$ condition. In this manner, Shannon's entropy
can be employed to build ordinary statistical mechanics and its
corresponding thermodynamics in which additivity and extensivity
are the backbone of all results. Some systems, such as those
including long range interactions, do not necessarily preserve the
additivity and extensivity properties
\cite{nn1,nn2,nn3,non0,non1,abe,1,2,fon}. These are generally the
systems described better by a power law distribution of
probabilities, namely $P_i^Q$ where $Q$ is a real parameter
\cite{pla}, instead of the ordinary $P_i$ distribution meaning
that other entropy measures are needed to describe these systems
\cite{non0,non1,SM1,SM2,pla}.

R\'{e}nyi ($\mathcal{S}$) and Tsallis ($S_T$) entropies are two
well-known of one-parameter generalized entropy defined as
\cite{nn3,non0,non1}

\begin{eqnarray}\label{rs}
&&\mathcal{S}=\frac{1}{\delta}\ln\sum_{i=1}^{W} P_i^{1-\delta},\\
&&S_T=\frac{1}{\delta}\sum_{i=1}^{W}\left(P_i^{1-\delta}-P_i\right),\nonumber
\end{eqnarray}

\noindent where $\delta\equiv1-Q$. Combining the above
one-parametric entropy measures with each other, we can find their
mutual relation \cite{nn3,non0,non1,non19,non20}

\begin{eqnarray}\label{reyn1}
\mathcal{S}=\frac{1}{\delta}\ln(1+\delta S_T).
\end{eqnarray}

There is also another generalized entropy measure, introduced
by Sharma and Mittal \cite{SM1,SM2}, indeed a two-parametric
entropy defined as \cite{pla,SM1,SM2,SM3,SM4,SM5}

\begin{eqnarray}\label{SME1}
S_{SM}=\frac{1}{1-r}\big((\sum_{i=1}^{W}
P_i^{1-\delta})^{\frac{1-r}{\delta}}-1\big),
\end{eqnarray}

\noindent where $r$ is a new free parameter. Some basic properties
of this entropy are addressed in
Refs.~\cite{pla,SM1,SM2,SM3,SM4,SM5} which show its compatibility
with various systems and indicate that it is a generalization of both
the R\'{e}nyi and Tsallis entropy. In fact, we can see that the
R\'{e}nyi and Tsallis entropies are recovered at the appropriate
limits of $r\rightarrow1$ and $r\rightarrow 1-\delta=Q$,
respectively \cite{pla,SM1,SM2,SM3,SM4,SM5}. Using Eqs.~(\ref{rs})
and~(\ref{SME1}), one can easily reach

\begin{eqnarray}\label{SME2}
S_{SM}=\frac{1}{R}\big((1+\delta S_T)^{\frac{R}{\delta}}-1\big),
\end{eqnarray}

\noindent where $R\equiv1-r$.

As we mentioned, systems including the long-range interactions are
better described by generalized entropies based on the power law
distributions of probability \cite{1,2,pla,fon}. Gravity is also a
long-range interaction which motivates physicist to use the
R\'{e}nyi and Tsallis generalized entropies in order to study the
gravitational and cosmological systems
\cite{non2,non15,non16,non17,non18,non19,non20,non21,non22,non23,
non3,non13,non4,non5,non6,non7,non8,non9,non10,non11,non12,non14,eb,eb1,3,4,5,6,7,8,9,10,11}.
Since $S_{SM}$ is the generalized form of both $\mathcal{S}$ and
$S_T$ \cite{pla,SM1,SM2,SM3,SM4,SM5}, we use $S_{SM}$ to build a
new HDE.

It has recently been argued that the Bekenstein-Hawking is a
proper candidate for the Tsallis entropy
\cite{abe,non2,non3,5,non18,non19,non20,non21,non22,non23}
allowing us to replace $S_T$ with $S_B$ in the above equation
which leads to

\begin{eqnarray}\label{SME3}
S_{SM}=\frac{1}{R}\big((1+\frac{\delta
A}{4})^{\frac{R}{\delta}}-1\big),
\end{eqnarray}

\noindent for the Sharma-Mittal entropy. In order to obtain this
result, we also used $S_B=\frac{A}{4}$, where $A$ is the horizon
area. For example, the Bekenstein-Hawking entropy is obtained by
using the Tsallis formalism in order to calculate the entropy of
black holes in loop quantum gravity \cite{5}. Thus, bearing
Eq.~(\ref{SME2}) in mind, we can say that Eq.~(\ref{SME3}) is in
fact the Sharma-Mittal entropy content of system.

\subsection*{Sharma-Mitall Holographic Dark Energy (SMHDE)}

Based on the holographic principle, the IR ($L$) and UV
($\Lambda$) cutoffs are in relation with the system horizon ($S$)
as \cite{HDE5,HDE4}

\begin{eqnarray}\label{coh}
\Lambda^4\propto\frac{S}{L^4},
\end{eqnarray}

\noindent In HDE hypothesis, the zero-point energy density
($\rho_\Lambda$) corresponding to the cut-off $\Lambda$ (
$\rho_\Lambda\sim\Lambda^4$), plays the role of the energy density
of dark energy ($\rho_D$) meaning that we have
$\rho_D\sim\Lambda^4$\cite{HDE5,HDE4}. Now, considering the Hubble
radius as the IR cutoff leading to $L\equiv
H^{-1}=\sqrt{\frac{A}{4\pi}}$, and by using Eqs.~(\ref{SME3})
and~(\ref{coh}), we reach at

\begin{eqnarray}\label{energydensity}
\rho_D=\frac{3C^2H^4}{8\pi
R}[(1+\frac{\delta\pi}{H^2})^{\frac{R}{\delta}}-1].
\end{eqnarray}

\noindent Here, $C^2$ is the unknown free parameter as usual, for
the energy density of SMHDE. The original HDE model is also
obtainable at the appropriate limit of $R\rightarrow\delta$. Here,
we consider a setup in which there is no interaction between
various components of cosmos, meaning that SMHDE obeys ordinary
conservation law, and thus

\begin{eqnarray}\label{emc1}
p_D=-(\frac{\dot{\rho}_D}{3H}+\rho_D)=-(\frac{\rho_D^\prime\dot{H}}{3H}+\rho_D),
\end{eqnarray}

\noindent where $\rho^\prime=\frac{d\rho_D}{dH}$, and dot denotes
derivative with respect to time.
\section{Universe evolution}

In a flat FRW universe, Friedmann equations are

\begin{eqnarray}\label{fe1}
&&H^2=\frac{8\pi}{3}(\rho_m+\rho_D),\\
&&H^2+\frac{2}{3}\dot{H}=\frac{-8\pi}{3}(p_D),\nonumber
\end{eqnarray}

\noindent where $\rho_m=\rho_0 a^{-3}$ and $\rho_0$ denote energy
density of matter fields, and its value at the current era
($a=1$), respectively. One can also obtain $p_D$ as a function of
$H$, by combining Eqs.~(\ref{energydensity}) and~(\ref{emc1}) with
the second Friedmann~(\ref{fe1}).

Now, defining $H^2(z)=E^2(z)H^2_0$, where $H_0$ is the current
value of the Hubble parameter, and
$\Omega_m=\frac{\rho_0}{\frac{3H_0^2}{8\pi}}$, Eq.~(\ref{fe1}) can
be rewritten as

\begin{eqnarray}\label{fe2}
E^2(z)&=&\Omega_m(1+z)^3\\&+&\frac{\frac{3C^2H_0^4E^4(z)}{8\pi
R}[(1+\frac{\delta\pi}{H_0^2E^2(z)})^{\frac{R}{\delta}}-1]}{\frac{3H_0^2}{8\pi}},\nonumber
\end{eqnarray}

\noindent which finally leads to

\begin{eqnarray}\label{fe6}
&&E^2(z)=\\&&\Omega_m(1+z)^3+\frac{(1-\Omega_m)E^4(z)}{[(1+\frac{\delta\pi}{H_0^2})^{\gamma}-1]}[(1+\frac{\delta\pi}{H_0^2E^2(z)})^{\gamma}-1]\nonumber,
\end{eqnarray}

\noindent where $\gamma\equiv\frac{R}{\delta}$. In obtaining the
above equation, we used the $E(z=0)=1$ condition which leads to

\begin{eqnarray}\label{fe4}
\frac{C^2H_0^2}{R}=\frac{1-\Omega_m}{[(1+\frac{\delta\pi}{H_0^2})^{\frac{R}{\delta}}-1]}.
\end{eqnarray}

\begin{figure}[ht]
\centering
\includegraphics[scale=0.4]{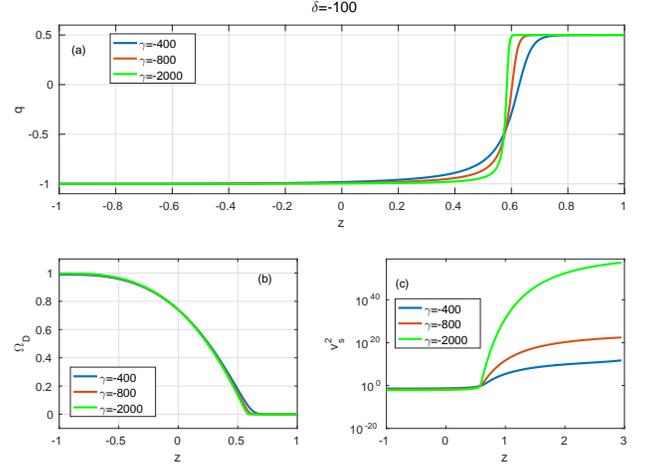}
\caption{$q$, $\Omega_D(z)$ and $v_{s}^{2}(z)$ versus $z$ for some
values of $\gamma$. Here, $\delta=-100$, $\Omega_m=0\cdot26$ and
$H_0=67\ (Km/s)/Mpc$.\label{delta}}
\end{figure}

\noindent The deceleration parameter is also evaluated as

\begin{eqnarray}\label{decp2}
q=-1+\frac{1+z}{E(z)}\frac{dE(z)}{dz}.
\end{eqnarray}

\noindent Moreover, since the density parameter of SMHDE is
defined as $\Omega_D(z)=\frac{\rho_D}{\frac{3H_0^2}{8\pi}}$, we
obtain

\begin{eqnarray}\label{omegaz}
\Omega_D(z)=\frac{(1-\Omega_m)E^4(z)}{[(1+\frac{\delta\pi}{H_0^2})^{\gamma}-1]}[(1+\frac{\delta\pi}{H_0^2E^2(z)})^{\gamma}-1],
\end{eqnarray}

\begin{figure}[ht]
\centering
\includegraphics[scale=0.4]{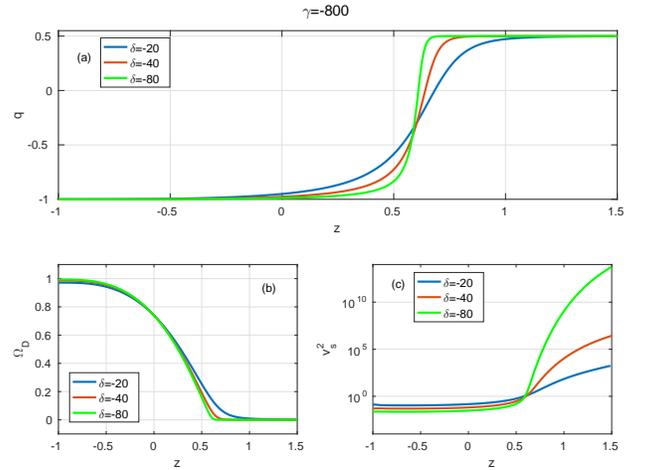}
\caption{$q$, $\Omega_D(z)$ and $v_{s}^{2}(z)$ versus $z$ for some
values of $\delta$. Here, $\gamma=-800$, $\Omega_m=0\cdot26$ and
$H_0=67\ (Km/s)/Mpc$.\label{gamma}}
\end{figure}

\noindent leading to $\Omega_D(z=0)=1-\Omega_m$ for current era, a
desired result. By inserting $\dot{H}=-4\pi p_D-\frac{3H^2}{2}$
and Eq.~(\ref{omegaz}) into Eq.~(\ref{emc1}), the pressure of
SMHDE is calculated as

\begin{eqnarray}\label{pf}
&&p_D=\frac{\rho_D-\frac{E(z)}{2}\frac{d\rho_D}{dE(z)}}{(\frac{4\pi}{3H_0^2E(z)})\frac{d\rho_D}{dE(z)}-1},
\end{eqnarray}

\noindent which can be used in order to study the stability of
system at classical level determined by the sign of the sound
velocity evaluated as

\begin{equation}\label{1}
v_{s}^{2}(z)=\frac{dp_D}{d\rho_D}=\frac{\frac{dp_D}{dz}}{\frac{d\rho_D}{dz}}=\frac{\frac{dp_D}{dE(z)}}{\frac{d\rho_D}{dE(z)}}.
\end{equation}

\noindent In fact, whenever $v_{s}^{2}(z)$ is positive, SMHDE is
stable.

In Figs.~(\ref{delta}) and~(\ref{gamma}), we have plotted $q$,
$\Omega_D(z)$ and $v_{s}^{2}(z)$ versus $z$ for some values of
$\gamma$ and $\delta$. From the panels ($a$) and ($b$) of both
figures, we see that there is a redshift ($z_d$) for which
$q\simeq\frac{1}{2}$ and $\Omega_D(z)\simeq0$, and its value is
increased at a fixed $\delta$ ($\gamma$), by increasing the value
of $\gamma$ ($\delta$). Moreover, panels ($c$) indicate that SMHDE
is stable for $z<z_d$ and changes in $v_{s}^{2}(z)$ will be more
relaxed by increasing the value of $\gamma$ ($\delta$) at a fixed
$\delta$ ($\gamma$). Indeed, $v_{s}^{2}(z)$ has a singularity at
$z=z_d$ and will be negative for $z>z_d$, meaning that such dark
energy candidate cannot remain stable in the matter dominated era,
and thus, the probable non-extensive features of spacetime cannot
affect and accelerate the universe expansion for $z>z_d$.
Therefore, our model may show better stability against the OHDE
constructed by considering Bekenstein-Hawking entropy and Hubble
horizon as its IR cutoff \cite{HDES}.

\begin{figure}[ht]
\centering
\includegraphics[scale=0.4]{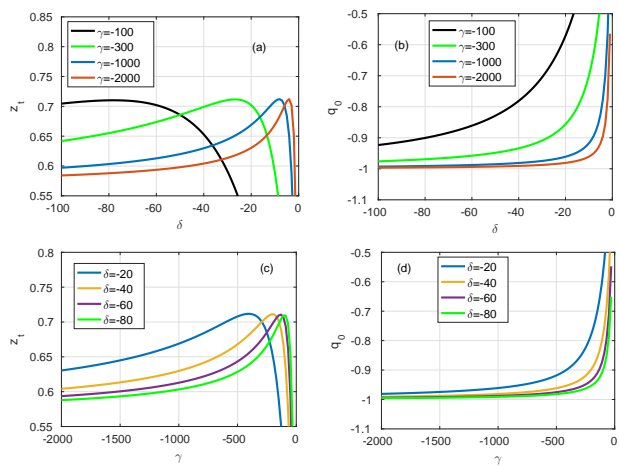}
\caption{$z_t$ and $q_0$ for some values of $\delta$ and $\gamma$,
whenever $\Omega_m=0\cdot26$ and $H_0=67\
(Km/s)/Mpc$.\label{trans}}
\end{figure}

The transition redshift ($z_t$), for which $q=0$, has also been
plotted in Fig.~(\ref{trans}) for some values of $\delta$ and
$\gamma$ (panels ($a$) and ($c$)) that lead to results compatible
with the $0\cdot55\leq z_t\leq0\cdot8$ range. Hence, this model
can theoretically produce proper values for $z_t$
\cite{Roos,RevH,wang,HDE5}. It is useful to note here that since
$\gamma$ and $\delta$ are negative, $R$ is positive in full
agreement with Eqs.~(\ref{fe4}),~(\ref{energydensity}) and thus the behavior of $\Omega_D(z)$. In
panels ($b$) and ($d$) of Fig.~(\ref{trans}), some values of
$\delta$ and $\gamma$ have been shown for them the current value
of the deceleration parameter ($q_0$) is within into the proper
range of $-0\cdot55\leq q_0\leq-1$ \cite{Roos}.

Finally, it is worthwhile mentioning that the total state
parameter defined as $w\equiv\frac{p_D}{\rho_D+\rho_m}$ can be
combined with Eq.~(\ref{decp2}) and~(\ref{fe1}) to reach at

\begin{eqnarray}\label{Eos}
w=\frac{2}{3}(q-\frac{1}{2}).
\end{eqnarray}

\noindent Comparing this equation with the behavior of $q$ and
$q_0$, plotted in Figs.~(\ref{delta}),~(\ref{gamma})
and~(\ref{trans}), one can see that $w$ has proper behavior during
the universe expansion in our model from matter dominated era
($w=0$), in which $q=\frac{1}{2}$, to the current era
($w\leq-\frac{2}{3}$), for which $q\leq-0\cdot55$ \cite{Roos}. In
addition, we can see that, at the $z\rightarrow-1$ limit since
$q\rightarrow-1$, our model predicts $w\rightarrow-1$. Finally, as
it is apparent from Fig.~(\ref{trans}), we should note here that
there are wide ranges for $\delta$ and $\gamma$ which can produce
desired results.


\section{Conclusion}

Attributing the Sharma-Mitall entropy to the horizon of flat FRW
universe, accepting the Bekenstein-Hawking entropy as the Tsallis
entropy, and by employing the holographic principle, we obtained a
HDE (SMHDE) in which Hubble horizon, in agreement with the
thermodynamics of flat FRW spacetime
\cite{Hay2,Hay22,Bak,CaiKim,Cai2}, plays the role of IR cutoff. In
addition, the evolution of universe filled by a pressureless
component and SMHDE, which do not have any mutual interaction, has
analytically been studied. Our approach shows that this model may
theoretically meet primary requirements to model the cosmos
expansion history. Therefore, our investigation offers that the
generalized statistical mechanics and its corresponding
thermodynamics have theoretically enough power to be in an
acceptable agreement with the behavior of cosmos motivating us to
more study the probable non-extensive aspects of spacetime.
\acknowledgments{The work of S. A. Moosavi has been supported
financially by Research Institute for Astronomy \& Astrophysics of
Maragha (RIAAM) under research project No. $1/5440-16$.}

\end{document}